\documentclass[prd,showpacs,preprintnumbers,amsmath,amssymb]{revtex4}

\usepackage{graphicx}
\usepackage{dcolumn}
\usepackage{bm}

\begin{document}

\title{Schwinger Pair Production at Finite Temperature in Scalar QED}
\author{Sang Pyo Kim}\email{sangkim@kunsan.ac.kr}
\affiliation{Department of Physics, Kunsan National University,
Kunsan 573-701, Korea} \affiliation{Asia
Pacific Center for Theoretical Physics, Pohang 790-784, Korea}

\author{Hyun Kyu Lee}\email{hyunkyu@hanyang.ac.kr}
\affiliation{Department of Physics and BK21 Division of Advanced
Research and Education in Physics, Hanyang University, Seoul
133-792}
\affiliation{Asia Pacific Center for Theoretical Physics,
Pohang 790-784, Korea}

\date{}

\begin{abstract}
In scalar QED we study the Schwinger pair production from an initial ensemble
of charged bosons when an electric field is turned on
for a finite period together with or without a constant magnetic field.
The scalar QED Hamiltonian depends on time through the electric
field, which causes the initial ensemble of bosons to evolve out of
equilibrium. Using the Liouville-von Neumann method for the density
operator and quantum states for each momentum
mode, we calculate the Schwinger pair-production rate at finite
temperature, which is the pair-production rate
from the vacuum times a thermal factor of the Bose-Einstein
distribution.
\end{abstract}

\pacs{12.20.-m, 13.40.-f, 11.10.Wx}

\maketitle

\section{Introduction}

Vacuum polarization and pair production has been known more than
seven decades ago \cite{Sauter}. Schwinger used the proper-time
method to find the exact one-loop effective action for a uniform
electromagnetic field \cite{Schwinger}. Since then vacuum
polarization and Schwinger pair production has been
investigated more or less from a theoretical view point
(for a review on recent
development of strong QED, see Ref. \cite{Dunne}). However, recently
the study of strong QED is more physically motivated than ever since
it will have many interesting physical applications in the near
future. In the near future terrestrial experiments, the x-ray free
electron lasers from Linac Coherent Light Source at SLAC and 
the TeV Energy Superconducting Linear Accelerator at DESY are expected to
achieve electromagnetic fields near the QED regime
\cite{Ringwald,slac,desy}. In astrophysics, most neutron stars are
expected to have magnetic fields comparable to the critical strength
$B_c = m^2c^2/e \hbar ~(4.4 \times 10^{13} G ~{\rm for ~electrons})$
and furthermore some magnetars are known to have magnetic fields a
few order above the critical strength. It is interesting that the
ratio of the pair-production rate in the
presence of parallel electric and magnetic fields to that in the
pure electric field is ${\cal N} (E, B)/{\cal N} (E,0) = (\pi qB/qE)
\times {\rm coth} (\pi qB /qE)$. This implies that the pair-production
rate is enhanced by the strong magnetic field, though the
leading behavior is still determined by the electric field as ${\cal
N} (E,0) = (qE)^{5/2} e^{- \pi m^2/qE} / (4 \pi^3 m)$. In these
compact stars pairs would be produced at non-zero temperature.

Another interesting issue from the view point of strong QED and
astrophysics is strange quark stars, hypothetical astrophysical objects.
Witten proposed that strange quark matter might be a stable state of nuclear
matter at ultrahigh density, which was numerically studied by Farhi and Jaffe
\cite{Witten}. Based on this hypothesis strange quark stars
were proposed as an alternative
to neutron stars \cite{Alcok}. The charge neutrality and chemical equilibrium
of strange quark stars would require a net amount of electrons, which are
bounded by the Coulomb attraction from the positive
core and form an electrosphere on the surface.
The Coulomb barrier from the electrosphere can generate an
electric field as large as $5 \times 10^{17}
V/cm$, which is stronger than the critical strength $E_c =
m^2 c^3/ e\hbar~(1.3 \times 10^{16} V/cm)$. The Coulomb barrier
may be a source of electron and positron pairs at finite temperature
\cite{Usov}. Therefore the Schwinger pair production at
finite temperature would be interesting for such astrophysical
objects.

It was Dittrich who first found the QED effective action at finite
temperature in the presence of a constant magnetic field
\cite{Dittrich}. Since then this has motivated the study of QED
at finite temperature under various configurations
of electromagnetic fields and thereby the Schwinger pair production
at finite temperature, which has been an issue of debate depending
on formalisms employed. The QED effective action for
a constant magnetic field was obtained at finite temperature
and density \cite{EPS}. Loewe and Rojas used the Schwinger
proper-time method to obtain the effective action in the presence
of both constant magnetic and electric fields and found an
imaginary part that has
contributions from vacuum as well as thermal fluctuations, leading
to a thermal enhancement of pair production at high temperature
\cite{Loewe-Rojas}. On the other hand, Elmfors and Skagerstam used
the real-time formalism to obtain the QED effective action in the presence of
both magnetic and electric fields, in which there is no imaginary part
and thereby no pair production \cite{Elmfors-Skagerstam}. Contrary
to this, Ganguly {\it et al} found in the finite temperature field theory
that the pair-production rate increases at high temperature
\cite{GKP}. Also in the functional Schr\"{o}dinger picture Hallin
and Liljenberg found the density matrix for QED when the electric
field is turned on only for a finite period and thereby calculated
the pair production at finite temperature when the system evolves
from an initial ensemble \cite{Hallin-Liljenberg}. Recently Gies used
the imaginary-time formalism to obtain the QED effective action in
the presence of both constant electric and magnetic fields, which
does not have any imaginary part at one-loop but nonzero imaginary
part at two-loop \cite{Gies99}.

The main purpose of this paper is to study the Schwinger pair
production from an initial ensemble of charged bosons in scalar QED
when an electric field is turned on only for a finite period
together with or without a constant magnetic field. A Sauter-type
electric field $E (t) = E_0 ~{\rm sech}^2 (t/\tau)$ provides an
analytical model \cite{Sauter32}, which acts effectively for $- \tau
\leq t \leq \tau$. The pair production from vacuum by this profile
of electric field extended in a finite time or space has been
studied
\cite{Narozhnyi-Nikishov,Ambjorn-Hughes,Balantekin-Fricke,Kim-Page,Kim-Page07,Dunne-Schubert,DWGS}.
Note that as the gauge potential $A_{\mu} = (0,0,0, -E_0 \tau \tanh
(t/\tau))$ in the time-dependent gauge depends on time explicitly,
so does the Hamiltonian. As the evolution operator $e^{- \beta H}$
in the Euclidean time is not the density operator, the
imaginary-time method based on the Matsubara frequency may not be
used directly for general time-dependent cases. In fact, the initial
ensemble evolves out of equilibrium due to the time-dependent
Hamiltonian, which requires a nonequilibrium quantum field theory.

In this paper we shall use a nonequilibrium quantum field theory
based on the Liouville-von Neumann equation, which was originally
introduced to find the density operator but led to exact quantum
states of time-dependent systems \cite{Lewis-Riesenfeld,Kim-Lee}.
The scalar QED Hamiltonian is quadratic in the field and momentum
and each momentum mode has a time-dependent frequency. The
time-dependent creation and annihilation operators that satisfy the
Liouville-von Neumann equation provide not only the density operator
but also exact quantum states. We find that the number of pairs
produced by the electric field is proportional to the initial
Bose-Einstein distribution. From this scalar QED model we may
conclude that the Schwinger pair-production rate by a time-dependent
electric field is enhanced by a thermal factor of the initial
Bose-Einstein distribution. The Liouville-von Neumann method can be
compared with other methods used to handle the scalar QED
Hamiltonian. Hallin and Liljenberg used the Schr\"{o}dinger picture
to find the wave functional of the Schr\"{o}dinger equation and the
density matrix \cite{Hallin-Liljenberg}, whereas the authors of
Refs. \cite{PSRSP,GGT} used the Heisenberg picture to find the time
evolution of the creation and annihilation operators for each
momentum of the field. It should be remarked that the time-dependent
creation and annihilation operators of the Liouville-von Neumann
equation provide not only the density operator but also exact
quantum states for each momentum.

The organization of this paper is as follows. In Sec. II, we explain
the canonical quantization method based on the Liouville-von Neumann
equation and apply it to scalar QED in the presence of an electric
field. In Sec. III, we
calculate the Schwinger pair-production rate at finite temperature
in the presence of Sauter-type electric field
together with or without a constant magnetic field.

\section{Canonical Quantization of Scalar QED}

Scalar QED for a charged boson with mass $m$ and charge $q$ under an
external electromagnetic field with the gauge field $A_{\mu}$ is described
by the Lagrangian [in units with $\hbar = c = 1$ and with metric
signature $(+, -, -, -)$]
\begin{equation}
L = \int d^3 x \Bigl[ \eta^{\mu \nu} (\nabla_{\mu} \Phi)^*
(\nabla_{\nu} \Phi) -  m^2 \Phi^* \Phi \Bigr], \label{lag}
\end{equation}
where the covariant derivative is given by
\begin{equation}
\nabla_{\mu} = \partial_{\mu} + i q A_{\mu}.
\end{equation}
For instance, the electric field in the $z$-direction may be described by the
time-dependent gauge potential $A_{\mu} = (0, 0, 0, A_z)$. Then the
Klein-Gordon equation from the Lagrangian (\ref{lag}) takes the form
\begin{equation}
 [\partial_t^2 - \partial_{\perp}^2 - (\partial_z + i q A_z(t))^2 + m^2 ] \Phi = 0.
\label{kg eq}
\end{equation}
The corresponding Hamiltonian is given by
\begin{equation}
H = \int d^3 x  \Bigl[\Pi^* \Pi + \Phi^* \Bigl(-
\partial_{\perp}^2 - (\partial_z + i q A_z)^2 + m^2 \Bigr) \Phi
\Bigr], \label{ham}
\end{equation}
where $\Pi = \dot{\Phi}^*$ and $\Pi^* = \dot{\Phi}$ are the
conjugate momenta.

Following Ref. \cite{Kim-Lee}, the fields
and momenta are decomposed into Fourier modes as
\begin{eqnarray}
\Phi ({\bf x}, t) &=& \int [d {\bf k}] \phi_{\bf k} e^{i {\bf k}
\cdot {\bf
x}},\nonumber\\
\Phi^* ({\bf x}, t) &=& \int [d {\bf k}] \phi^*_{\bf k} e^{- i {\bf
k} \cdot {\bf
x}},
\end{eqnarray}
where $[d {\bf k}] = d^3 {\bf k}/(2 \pi)^3$ and
\begin{eqnarray}
\Pi ({\bf x}, t) &=&  \int [d {\bf k}]
\dot{\phi}^*_{\bf k} e^{-i {\bf k} \cdot {\bf x}} = \int [d {\bf k}]
\pi_{\bf k} e^{- i {\bf k} \cdot {\bf x}},\nonumber \\
\Pi^* ({\bf x}, t) &=&  \int [d {\bf k}]
\dot{\phi}_{\bf k} e^{i {\bf k} \cdot {\bf x}} = \int [d {\bf k}]
\pi^*_{\bf k} e^{i {\bf k} \cdot {\bf x}}.
\end{eqnarray}
The Hamiltonian is then given by
\begin{eqnarray}
H = \int [d {\bf k}]  \Bigl[\pi_{\bf k}^* \pi_{\bf k} + \omega_{\bf k}^2
(t) \phi_{\bf k}^* \phi_{\bf k} \Bigr], \label{ham dec}
\end{eqnarray}
where
\begin{equation}
\omega^2_{\bf k} (t)  = (k_z + q A_z(t))^2 + {\bf k}_{\perp}^2 +
m^2. \label{omega}
\end{equation}
Here, $\phi_{\bf k}, \phi^*_{\bf k}$ and $\pi_{\bf k}, \pi^*_{\bf k}$ are
Schr\"{o}dinger operators. Each Fourier mode obeys the equation
\begin{eqnarray}
[\partial_t^2 + \omega^2_{\bf k} (t) ] \phi_{\bf k} (t) = 0.
\label{m eq}
\end{eqnarray}

Upon canonical quantization, we impose
the equal-time commutation relation $[\Phi ({\bf x}', t), \Pi
({\bf x}, t)] = i \delta ({\bf x}' - {\bf x})$ and
$[\Phi^* ({\bf x}', t), \Pi^*
({\bf x}, t)] = i \delta ({\bf x}' - {\bf x})$, from which follow
the commutation relations
\begin{eqnarray}
\bigl[ \phi_{ {\bf k}' }, \pi_{\bf k} \bigr] = i (2 \pi)^3
\delta ( {\bf k}' - {\bf k} ), \nonumber\\
\bigl[ \phi^*_{ {\bf k}' }, \pi^*_{\bf k} \bigr] = i (2 \pi)^3
\delta ({\bf k}' - {\bf k} ),
\end{eqnarray}
but all the other commutators vanish.
We may introduce the time-dependent annihilation and creation operators
to quantize the fields as
\begin{eqnarray}
\Phi ({\bf x}, t) &=& \int [d {\bf k}] \Bigl[ \varphi_{\bf k} (t) a_{\bf k}
(t) + \varphi^*_{\bf k} (t) b^{\dagger}_{\bf k} (t) \Bigr]  e^{i {\bf k}
\cdot {\bf
x}},\nonumber\\
\Phi^* ({\bf x}, t) &=& \int [d {\bf k}] \Bigl[ \varphi_{\bf k} (t) b_{\bf k}
(t) + \varphi^*_{\bf k} (t) a^{\dagger}_{\bf k} (t) \Bigr]  e^{- i {\bf k}.
\cdot {\bf
x}}, \label{phi quan}
\end{eqnarray}
and similarly the momenta as
\begin{eqnarray}
\Pi ({\bf x}, t) &=& \int [d {\bf k}] \Bigl[ \dot{\varphi}^*_{\bf k} (t)
a^{\dagger}_{\bf k} (t) + \dot{\varphi}_{\bf k} (t) b_{\bf k} (t) \Bigr]  e^{-i {\bf k}
\cdot {\bf
x}},\nonumber\\
\Pi^* ({\bf x}, t) &=& \int [d {\bf k}] \Bigl[ \dot{\varphi}^*_{\bf k} (t)
b^{\dagger}_{\bf k}(t) + \dot{\varphi}_{\bf k} (t) a_{\bf k} (t) \Bigr]  e^{i {\bf k}.
\cdot {\bf
x}}. \label{pi quan}
\end{eqnarray}
A few comments are in order. First, note that $a_{\bf k}$
and $a^{\dagger}_{\bf k}$ correspond to the particle and
$b_{\bf k}$ and $b^{\dagger}_{\bf k}$ to the antiparticle and that
$\Phi$ annihilates particles but creates antiparticles. Second,
it can be shown that the field quantization includes static fields as
a specific case
by choosing $\varphi_{\bf k} = e^{-i \omega_{\bf k} t}/\sqrt{2 \omega_{\bf k}}$.
Finally, by imposing the Wronskian conditions
\begin{equation}
\dot{\varphi}^*_{\bf k} (t) \varphi_{\bf k} (t) - \varphi^*_{\bf k}
(t) \dot{\varphi}_{\bf k} (t) = i, \label{wron}
\end{equation}
we have the standard equal-time commutation relations
\begin{eqnarray}
\bigl[ a_{ {\bf k}' } (t), a^{\dagger}_{\bf k} (t) \bigr]
= \delta ( {\bf k}' - {\bf k} ), \quad
\bigl[ b_{ {\bf k}' } (t), b^{\dagger}_{\bf k} (t) \bigr] = \delta ({\bf k}' - {\bf
k} ).
\end{eqnarray}
All the other commutators vanish.

By inverting the Fourier modes in Eqs. (\ref{phi quan})
and (\ref{pi quan}), we find
two independent sets of time-dependent annihilation and
creation operators \cite{Kim-Lee}
\begin{eqnarray}
a_{\bf k} (t) &=& i \bigl[ \varphi^*_{\bf k} (t) \pi^*_{\bf k} -
\dot{\varphi}^*_{\bf k} (t) \phi_{\bf k} \bigr], \nonumber\\
a^{\dagger}_{\bf k} (t) &=& - i \bigl[ \varphi_{\bf k} (t) \pi_{\bf k} -
\dot{\varphi}_{\bf k} (t) \phi^*_{\bf k} \bigr], \label{a op}
\end{eqnarray}
and
\begin{eqnarray}
b_{\bf k} (t) &=& i \bigl[ \varphi^*_{\bf k} (t) \pi_{\bf k} -
\dot{\varphi}^*_{\bf k} (t) \phi^*_{\bf k} \bigr], \nonumber\\
b^{\dagger}_{\bf k} (t) &=& - i \bigl[ \varphi_{\bf k} (t) \pi^*_{\bf k}
- \dot{\varphi}_{\bf k} (t) \phi_{\bf k} \bigr]. \label{b op}
\end{eqnarray}
It is remarkable that when $\varphi_{\bf k}$ satisfies the mode equation (\ref{m eq}),
the time-dependent operators (\ref{a op}) and (\ref{b op})
satisfy the Liouville-von Neumann equation, for instance,
\begin{equation}
i \frac{\partial a_{\bf k} (t)}{\partial t} + [ a_{\bf k} (t),
H (t)] = 0.
\end{equation}
This implies that we may define the density operator for particles
\begin{eqnarray}
\rho_{a_{\bf k}} (t) = \frac{1}{Z_{\bf k}} \exp \Bigl[- \frac{\omega_{\bf k}}{kT}
\Bigl( a^{\dagger}_{\bf k} (t) a_{\bf k} (t) + \frac{1}{2} \Bigr) \Bigr], \label{a den}
\end{eqnarray}
and for antiparticles
\begin{eqnarray}
\rho_{b_{\bf k}} (t) = \frac{1}{Z_{\bf k}} \exp \Bigl[- \frac{\omega_{\bf k}}{kT}
\Bigl( b^{\dagger}_{\bf k} (t) b_{\bf k} (t) + \frac{1}{2} \Bigr) \Bigr]. \label{b den}
\end{eqnarray}
Here, $Z_{\bf k} = {\rm Tr} (e^{- \omega_{\bf k} ( a^{\dagger}_{\bf k} (t)
a_{\bf k} (t) + 1/2)/kT}) = {\rm Tr} (e^{- \omega_{\bf k} ( b^{\dagger}_{\bf k} (t)
b_{\bf k} (t) + 1/2)/kT})$. We note that $T$ is the temperature of an
initial ensemble of the heat bath of particles before the onset of electric
fields and that the number of produced pairs should be small not to
change the temperature of heat bath.

We now turn to pair production by electric fields.
The mode-decomposed Hamiltonian (\ref{ham dec}) is
a collection of time-dependent harmonic oscillators.
The equation of motion (\ref{m eq}), in the language of
quantum mechanics, becomes the
super-barrier transmission over a potential barrier
in the time dimension. Pair production may be explained in such a
way that the positive-frequency solution at the past
infinity scatters not only to the
positive-frequency solution but also to the
negative-frequency solution at the future infinity.
Then the magnitude square of the ratio of the coefficient
of the negative-frequency solution at the future infinity to the
coefficient of the positive-frequency solution at the past infinity
is the pair-production rate \cite{Parker,DeWitt}.
However, we shall employ the Liouville-von Neumann method, in which we first find
the operators known as the Lewis-Riesenfeld
or invariant operators that satisfy the Liouville-von Neumann equation
\cite{Lewis-Riesenfeld} and then construct not only
the exact quantum states but also the
density operator. The time-dependent vacuum is defined as
\begin{eqnarray}
a_{\bf k} (t) \vert 0; t \rangle = b_{\bf k} (t) \vert 0; t \rangle = 0,
\quad (\forall {\bf k})
\label{vac}
\end{eqnarray}
and the multi-particle and antiparticle states as
\begin{eqnarray}
\vert n_{{\bf k}_1} \cdots; n_{{\bf k}_2} \cdots; t \rangle =
\frac{a^{\dagger n_1}_{{\bf k}_1} (t)}{\sqrt{n_1!}} \cdots
\frac{b^{\dagger n_2}_{{\bf k}_2} (t)}{\sqrt{n_2!}} \cdots \vert 0; t \rangle.
\label{num st}
\end{eqnarray}
The time-dependent vacuum (\ref{vac}) is the zero-particle
and antiparticle state of the number operators
\begin{eqnarray}
N_{a_{\bf k}} (t) = a^{\dagger}_{\bf k} (t) a_{\bf k} (t), \quad
N_{b_{\bf k}} (t) = b^{\dagger}_{\bf k} (t) b_{\bf k} (t).
\end{eqnarray}
It should be remarked that the Fock states are exact quantum states of the
time-dependent Hamiltonian up to some time-dependent phase factors
{\it a la} the Liouville-von Neumann method
\cite{Lewis-Riesenfeld,Kim-Lee}.

To compare with the standard quantization of field, we consider the limiting
case of no electromagnetic field. Then our definition of the
time-dependent annihilation operator differs from the standard
one by an overall time-dependent phase factor:
\begin{eqnarray}
a_{\bf k} (t) = e^{ i \omega_{\bf k} t} a_{\bf k}
\end{eqnarray}
for the choice of $\varphi_{\bf k} = e^{- i \omega_{\bf k} t}/
\sqrt{2 \omega_{\bf k}}$.
The phase factor, which is necessary for the Liouville-von Neumann equation,
does not change the number operators
$N_{a_{\bf k}} (t) = N_{a_{\bf k}}$ and $N_{b_{\bf k}} (t) = N_{b_{\bf k}}$.
In this sense
the Liouville-von Neumann method includes
the standard quantization as a specific case. The method turns out to be
convenient and powerful for time-dependent quantum fields such as scalar QED in
the presence of time-dependent electromagnetic fields. The number of
pairs with a certain momentum ${\bf k}$ produced by an external field is
the number of particles or antiparticles of the out-vacuum at $t =  \infty$
contained in the in-vacuum at $t = - \infty$  \cite{Parker,DeWitt}
\begin{eqnarray}
n_{\bf k} = \langle 0; {\rm in} \vert N_{a_{\bf k}} ( t = \infty)
\vert 0; {\rm in} \rangle = \langle 0; {\rm in} \vert
N_{b_{\bf k}} ( t = \infty) \vert 0; t = {\rm in} \rangle.
\end{eqnarray}

\section{Pair Production at Finite Temperature}

We first study the Schwinger pair production by a
uniform field $E(t) = E_0 {\rm sech}^2 (t/\tau)$ without a magnetic field
at finite temperature and then study the case with a constant magnetic field $B$.
When ${\bf E} \cdot {\bf B} \neq 0$ we can find a Lorentz frame where the
electric and magnetic fields are parallel to each other,
for instance, in the $z$-direction. We choose is a Sauter-type gauge potential
$A_{\mu} = (0, - By/2, Bx/2, - E_0 \tau \tanh (t/\tau))$. It is shown that
pair production by the Sauter-type potential
is characterized by two parameters \cite{Kim-Page07}
\begin{eqnarray}
\epsilon_t = \frac{m}{qE_0 \tau}, \quad \delta_t =
\frac{qE_0}{\pi m^2}.
\end{eqnarray}
The pair production is energetically favored for $\epsilon_t
< 1$ and the adiabaticity parameter $\delta_t$ determines the pair-production rate.

\subsection{Pure Electric Field}

The mode equation for the auxiliary field variable takes the form
\begin{equation}
\ddot{\varphi}_{\bf k} (t) + \omega_{\bf k}^2 (t) \varphi_{\bf k} (t) = 0,
\label{aux eq}
\end{equation}
where
\begin{eqnarray}
\omega_{\bf k}^2 (t) = \Bigl(k_z - qE_0 \tau \tanh (\frac{t}{\tau}) \Bigr)^2
+ {\bf k}^2_{\perp} +m^2.
\end{eqnarray}
Each mode has two asymptotic frequencies at $t = \mp \infty$
\begin{eqnarray}
\omega^{(-)}_{\bf k} = \sqrt{(k_z + qE_0 \tau)^2
+ {\bf k}^2_{\perp} +m^2}, \nonumber \\
\omega^{(+)}_{\bf k} = \sqrt{(k_z - qE_0 \tau)^2
+ {\bf k}^2_{\perp} +m^2}.
\end{eqnarray}
The solution with the asymptotic form
\begin{eqnarray}
\varphi^{\rm in}_{\bf k} (t) = \frac{
e^{- i \omega^{(-)}_{\bf k} t}}{\sqrt{2 \omega^{(-)}_{\bf k}}} \label{asym fr-}
\end{eqnarray}
at $t = - \infty$ is given by \cite{Narozhnyi-Nikishov,Ambjorn-Hughes}
\begin{eqnarray}
\varphi_{\bf k} (t) = \frac{1}{\sqrt{2 \omega^{(-)}_{\bf k}
e^{\pi \tau \omega^{(-)}_{\bf k}}}} (1 - z)^{1/2 + i \lambda}
z^{- i \tau \omega^{(-)}_{\bf k}/2} F(\alpha_{\bf k},
\beta_{\bf k}; \gamma_{\bf k}; z),
\label{sol}
\end{eqnarray}
where $F(\alpha_{\bf k}, \beta_{\bf k}; \gamma_{\bf k}; z)$ is the hypergeometric function and
\begin{eqnarray}
z &=& - 2 e^{2 t/\tau},\nonumber \\
\lambda &=& \sqrt{(qE_0 \tau^2)^2 - \frac{1}{4}}
\end{eqnarray}
and
\begin{eqnarray}
\alpha_{\bf k} &=& \frac{1}{2} - \frac{i}{2} \bigl(\tau \omega^{(-)}_{\bf k}
- \tau \omega^{(+)}_{\bf k}  - 2 \lambda \bigr), \nonumber \\
\beta_{\bf k} &=& \frac{1}{2} - \frac{i}{2} \bigl(\tau \omega^{(-)}_{\bf k}
+ \tau \omega^{(+)}_{\bf k}  - 2 \lambda \bigr), \nonumber \\
\gamma_{\bf k} &=& 1 - i \tau \omega^{(-)}_{\bf k}.
\end{eqnarray}
Note that the solution (\ref{sol}) satisfies the
Wronskian condition (\ref{wron}) for any time. At the other asymptotic limit
$t = \infty$, the solution becomes
\begin{eqnarray}
\varphi^{(+)}_{\bf k} (t) = A_{\bf k} \varphi^{\rm out}_{\bf k} (t)
+ B_{\bf k} \varphi^{\rm out *}_{\bf k} (t),  \label{asym fr+}
\end{eqnarray}
where
\begin{eqnarray}
\varphi^{\rm out}_{\bf k} (t) = \frac{e^{- i \omega^{(+)}_{\bf k} t}}{\sqrt{2 \omega^{(+)}_{\bf k}}}, \label{asym out}
\end{eqnarray}
and
\begin{eqnarray}
A_{\bf k} &=& 2^{- i \tau \omega^{(+)}_{\bf k}}\sqrt{
\frac{ \omega^{(+)}_{\bf k}}{\omega^{(-)}_{\bf k}}}
\frac{\Gamma (\gamma_{\bf k}) \Gamma(\beta_{\bf k} - \alpha_{\bf k})}{
\Gamma (\beta_{\bf k})
\Gamma (\gamma_{\bf k} - \alpha_{\bf k})}, \nonumber \\
B_{\bf k} &=& 2^{i \tau \omega^{(+)}_{\bf k}}
\sqrt{\frac{\omega^{(+)}_{\bf k}}{\omega^{(-)}_{\bf k}}}
\frac{\Gamma (\gamma_{\bf k}) \Gamma(\alpha_{\bf k}
- \beta_{\bf k})}{\Gamma (\alpha_{\bf k})
\Gamma (\gamma_{\bf k} - \beta_{\bf k})}.
\end{eqnarray}
As the asymptotic form (\ref{asym fr+}) is approximately valid
for $t \geq \tau$, the analysis below is a good approximation after
$\tau$ and exact at $t = \infty$.

The operators in Eqs. (\ref{a op}) and (\ref{b op}) obtained
by substituting the asymptotic solution (\ref{asym fr-}) define the
time-dependent annihilation and creation operators $a^{\rm in}_{\bf k},
a^{\dagger{\rm in}}_{\bf k}$ and $b^{\rm in}_{\bf k},
b^{\dagger{\rm in}}_{\bf k}$ for the in-vacuum while those
obtained by the asymptotic solution (\ref{asym out}) define the time-dependent
annihilation and creation operators $a^{\rm out}_{\bf k},
a^{\dagger{\rm out}}_{\bf k}$ and $b^{\rm out}_{\bf k},
b^{\dagger{\rm out}}_{\bf k}$ for the out-vacuum. The mode solution starting
from the initial asymptotic solution (\ref{asym fr-}) evolves to
another asymptotic form (\ref{asym fr+}). This leads to a Bogoliubov
transformation between the in-vacuum operators and the out-vacuum operators
\begin{eqnarray}
a^{\rm out}_{\bf k} &=& A_{\bf k} a^{\rm in}_{\bf k} +
B^*_{\bf k} b^{\dagger{\rm in}}_{\bf k}, \nonumber\\
b^{\rm out}_{\bf k} &=& A_{\bf k} b^{\rm in}_{\bf k} +
B^*_{\bf k} a^{\dagger{\rm in}}_{\bf k},
\end{eqnarray}
whose coefficients satisfy the relation
\begin{eqnarray}
|A_{\bf k}|^2 - |B_{\bf k}|^2 = 1.
\end{eqnarray}

The in-vacuum is the superposition of the out-vacuum particles.
The number operator of the in-vacuum is $ N^{\rm in}_{a_{\bf k}}
= a^{\dagger{\rm in}}_{\bf k} a^{\rm in}_{\bf k}$ for particles
and  $ N^{\rm in}_{b_{\bf k}}
= b^{\dagger{\rm in}}_{\bf k} b^{\rm in}_{\bf k}$ for the antiparticles.
Similarly, the particle number operator of the out-vacuum is
$ N^{\rm out}_{a_{\bf k}}
= a^{\dagger{\rm out}}_{\bf k} a^{\rm out}_{\bf k}$ and the antiparticle
number operator is $ N^{\rm out}_{b_{\bf k}}
= b^{\dagger{\rm out}}_{\bf k} b^{\rm out}_{\bf k}$. Hence the
number of created pairs by the electric field from the vacuum is
\begin{eqnarray}
n_{\bf k} = \langle 0; {\rm in} \vert N^{\rm out}_{a_{\bf k}}
\vert 0; {\rm in} \rangle = \langle 0; {\rm in} \vert N^{\rm out}_{b_{\bf k}}
\vert 0; {\rm in} \rangle = |B_{\bf k}|^2,
\end{eqnarray}
where
\begin{eqnarray}
|B_{\bf k}|^2 = \frac{\cosh[\pi(2\lambda + \tau \omega_{\bf k}^{(+)}
- \tau \omega_{\bf k}^{(-)})/2]
\cosh[\pi(2\lambda + \tau \omega_{\bf k}^{(-)}
- \tau \omega_{\bf k}^{(+)})/2]}{\cosh[\pi(2\lambda + \tau \omega_{\bf k}^{(-)}
+ \tau \omega_{\bf k}^{(+)})/2]
\cosh[\pi(2\lambda - \tau \omega_{\bf k}^{(-)}
- \tau \omega_{\bf k}^{(+)})/2]}. \label{pair}
\end{eqnarray}
The number of pairs (\ref{pair}) approximately is
\begin{eqnarray}
|B_{\bf k}|^2 \approx e^{- \pi (\tau \omega_{\bf k}^{(-)}
+ \tau \omega_{\bf k}^{(+)} - \lambda)} = e^{- S_{\bf k}^{(0)}},
\end{eqnarray}
where $S_{\bf k}^{(0)}$ is the leading-order contribution of the
WKB instanton action in scalar QED \cite{Kim-Page07}.

The density operator (\ref{a den}) takes the asymptotic form for particles
\begin{eqnarray}
\rho^{\rm in}_{a_{\bf k}} = \frac{1}{Z_{\bf k}} \exp \Bigl[-
\frac{\omega^{(-)}_{\bf k}}{kT}
\Bigl( N^{\rm in}_{a_{\bf k}} + \frac{1}{2} \Bigr) \Bigr],
\end{eqnarray}
and the density operator (\ref{b den}) for antiparticles
\begin{eqnarray}
\rho^{\rm in}_{b_{\bf k}} = \frac{1}{Z_{\bf k}} \exp \Bigl[-
\frac{\omega^{(-)}_{\bf k}}{kT}
\Bigl( N^{\rm in}_{b_{\bf k}} + \frac{1}{2} \Bigr) \Bigr].
\end{eqnarray}
Hence the initial ensemble of particles has the Bose-Einstein distribution
\begin{eqnarray}
f_{\bf k} (T) = {\rm Tr} \Bigl(\rho^{\rm in}_{\bf k} N^{\rm in}_{a_{\bf k}}
\Bigr)
= \frac{1}{e^{\omega^{(-)}_{\bf k}/kT} -1}.
\end{eqnarray}
Finally, the number of pairs produced by the electric field
from the initial thermal ensemble is
\begin{eqnarray}
n_{\bf k} (E, T) = {\rm Tr} \Bigl(\rho^{\rm in}_{\bf k} N^{\rm out}_{a_{\bf k}}
\Bigr) - f_{\bf k}
= |B_{\bf k} (E)|^2( 2 f_{\bf k} (T)  + 1).
\label{sch fin}
\end{eqnarray}
The thermal factor can be written as $2 f_{\bf k} +1 = {\rm coth}
(\omega_{\bf k}^{(-)}/2 kT)$.
We have thus found that the Schwinger pair-production rate at finite temperature
is enhanced by the thermal factor $f_{\bf k}$. In the zero-temperature
limit we have $f_{\bf k} = 0$ and recover the Schwinger pair-production
rate from the vacuum. The Schwinger pair-production
rate per unit volume and per unit time is given by summing over all momenta
\begin{eqnarray}
{\cal N} (E, T) = \int [d{\bf k}] n_{\bf k} (E, T).
\label{tot rate}
\end{eqnarray}
As the solution (\ref{sol}) approximately approaches
the asymptotic form (\ref{asym fr-}) for $t \approx - \tau$ and
(\ref{asym fr+}) for $t \approx \tau$, the
pair-production rates (\ref{sch fin}) and (\ref{tot rate}) are
approximately valid after $t = \tau$.

In the asymptotic region where the electric field is turned off,
the out-vacuum is well-defined and the excited states (\ref{num st})
correspond to a number of particles and/or antiparticles.
In this region the pair-production rate (\ref{sch fin})
or (\ref{tot rate}) has the interpretation of created pairs. Now
a question may be raised how to interpret Eqs. (\ref{sch fin}) or
(\ref{tot rate}) while the electric field is acting.
In terms of the in-vacuum operators we may express
the time-dependent annihilation operator at any time as
\begin{eqnarray}
a_{\bf k} (t) &=& A_{\bf k} (t) a^{\rm in}_{\bf k} + B^*_{\bf k}
(t) b_{\bf k}^{{\rm in}\dagger}, \nonumber\\
b_{\bf k} (t) &=& A_{\bf k} (t) b^{\rm in}_{\bf k} + B^*_{\bf k}
(t) a_{\bf k}^{{\rm in}\dagger},
\end{eqnarray}
where
\begin{eqnarray}
A_{\bf k} (t) &=& i \Bigl(\varphi_{\bf k} (t) \dot{\varphi}^{\rm in}_{\bf k}
- \dot{\varphi}^*_{\bf k} (t) \varphi^{\rm in}_{\bf k} \Bigr), \nonumber\\
B_{\bf k} (t) &=& i \Bigl(\varphi^*_{\bf k} (t) \dot{\varphi}^{{\rm in}*}_{\bf k}
- \dot{\varphi}^*_{\bf k} (t) \varphi^{{\rm in}*}_{\bf k} \Bigr).
\end{eqnarray}
Here $\varphi^{\rm in}_{\bf k}$ is the solution at $t = - \infty$. Then
we can obtain
\begin{eqnarray}
n_{\bf k} (E, T; t) = {\rm Tr} \Bigl(\rho^{\rm in}_{\bf k} N_{a_{\bf k}} (t)
\Bigr) - f_{\bf k}
= |B_{\bf k} (E; t)|^2( 2 f_{\bf k} (T)  + 1).
\label{sch fin t}
\end{eqnarray}
It is tempting to interpret (\ref{sch fin t}) as
the number of particles at time $t$.
However, the concept of particle can have a meaning only
in the adiabatic limit where
\begin{eqnarray}
\varphi^{\rm ad}_{\bf k} (t) = \frac{e^{- i \int
\omega_{\bf k} (t)}}{\sqrt{2 \omega_{\bf k} (t)}}
\end{eqnarray}
is an approximate solution to the mode equation (\ref{aux eq}).
The solution is approximately good for a large momentum
since $\omega_{\bf k} \gg |\dot{\omega}_{\bf k}|,
|\ddot{\omega}_{\bf k}|$. For a small momentum the change of
the gauge potential and the electric field should be small to
guarantee the validity of
the adiabatic solution. In the adiabatic case the solution with the
initial asymptotic form (\ref{asym fr-}) evolves approximately
to the form
\begin{eqnarray}
\varphi_{\bf k} (t) \approx A_{\bf k} (t) \varphi^{\rm ad}_{\bf k} (t)
+ B_{\bf k} (t) \varphi^{{\rm ad}*}_{\bf k} (t).
\end{eqnarray}
In the other case where the adiabatic approximation
breaks down, we cannot have a concept of particle for
such momentum modes but Eq. (\ref{sch fin t})
may be interpreted as a distribution.

\subsection{Electric and Magnetic Fields}

We now consider the case of a uniform field $E(t) = E_0 {\rm sech}^2 (t/\tau)$
together with a parallel magnetic field $B$.
The gauge potential $A_{\mu} = (0, - By/2, Bx/2, - E_0 \tau \tanh (t/\tau))$.
The mode equation takes the form
\begin{equation}
\ddot{\varphi}_{k} (t, x, y) + \omega_{k}^2 (t) \varphi_{k} (t, x, y) = 0,
\end{equation}
where
\begin{eqnarray}
\omega^2_{k} (t) = H_{\perp} +
\Bigl(k_z - qE_0 \tau \tanh (\frac{t}{\tau}) \Bigr)^2 + m^2.
\end{eqnarray}
Here $H_{\perp}$ describes the two-dimensional motion
transverse to the magnetic field
\begin{eqnarray}
H_{\perp} &=& - \Bigl(\partial_x - i \frac{qB}{2} y \Bigr)^2 -
\Bigl(\partial_y + i \frac{qB}{2} x \Bigr)^2 \nonumber \\
&=& - (\partial_x^2 + \partial_y^2) + \Bigl(\frac{qB}{2} \Bigr)^2
( x^2 + y^2) - \frac{qB}{2} L_z. \label{tran}
\end{eqnarray}
It has the eigenvalue
\begin{eqnarray}
H_{\perp} \varphi_{n} (x,y) = qB(2n+1) \varphi_{n} (x, y),
\end{eqnarray}
whose orthonormal eigenfunctions were given in Ref. \cite{Johnson-Lippmann}.
Then the mode $\varphi_{k} (t, x, y)= \varphi_{nk} (t)
\varphi_{n}(x,y)$ satisfies
\begin{equation}
\ddot{\varphi}_{nk} (t) + \omega_{nk}^2 (t) \varphi_{nk} (t) = 0,
\end{equation}
where
\begin{eqnarray}
\omega_{nk}^2 = \Bigl(k_z - qE_0 \tau \tanh (\frac{t}{\tau}) \Bigr)^2
+ qB(2n+1) + m^2.
\end{eqnarray}
The two asymptotic frequencies at $t = \mp \infty$ are given by
\begin{eqnarray}
\omega^{(-)}_{nk} = \sqrt{(k_z + qE_0 \tau)^2
+ qB(2n+1) +m^2}, \nonumber\\
\omega^{(+)}_{nk} = \sqrt{(k_z - qE_0 \tau)^2
+ qB(2n+1) +m^2}.
\end{eqnarray}
The mode has the solution of the same form as Eq. (\ref{sol})
\begin{eqnarray}
\varphi_{nk} (t) = \frac{1}{\sqrt{2 \omega^{(-)}_{nk}
e^{\pi \tau \omega^{(-)}_{nk}}}} (1 - z)^{1/2 + i \lambda}
z^{- i \tau \omega^{(-)}_{nk}/2} F(\alpha_{nk}, \beta_{nk};
\gamma_{nk}; z),
\end{eqnarray}
with parameters replaced by
\begin{eqnarray}
\alpha_{nk} &=& \frac{1}{2} - \frac{i}{2} \bigl(\tau \omega^{(-)}_{nk}
- \tau \omega^{(+)}_{nk}  - 2 \lambda \bigr), \nonumber\\
\beta_{nk} &=& \frac{1}{2} - \frac{i}{2} \bigl(\tau \omega^{(-)}_{nk}
+ \tau \omega^{(+)}_{nk}  - 2 \lambda \bigr), \nonumber \\
\gamma_{nk} &=& 1 - i \tau \omega^{(-)}_{nk}.
\end{eqnarray}
The Bogoliubov transformations for particles and antiparticles are given by
\begin{eqnarray}
a^{\rm out}_{nk} &=& A_{nk} a^{\rm in}_{nk} +
B^*_{nk} b^{\dagger{\rm in}}_{nk}, \nonumber\\
b^{\rm out}_{nk} &=& A_{nk} b^{\rm in}_{nk} +
B^*_{nk} a^{\dagger{\rm in}}_{nk},
\end{eqnarray}
where
\begin{eqnarray}
A_{nk} &=& 2^{- i \tau \omega^{(+)}_{nk}} \sqrt{
\frac{ \omega^{(+)}_{nk}}{\omega^{(-)}_{nk}}}
\frac{\Gamma (\gamma_{nk}) \Gamma(\beta_{nk}
- \alpha_{nk})}{\Gamma (\beta_{nk})
\Gamma (\gamma_{nk} - \alpha_{nk})}, \nonumber\\
B_{nk} &=& 2^{i \tau \omega^{(+)}_{nk}}
\sqrt{\frac{\omega^{(+)}_{nk}}{\omega^{(-)}_{nk}}}
\frac{\Gamma (\gamma_{nk}) \Gamma(\alpha_{nk}
- \beta_{nk})}{\Gamma (
\alpha_{nk})
\Gamma (\gamma_{nk} - \beta_{nk})}.
\end{eqnarray}
where
\begin{eqnarray}
f_{nk} (T) = {\rm Tr} \Bigl(\rho^{\rm in}_{nk} N^{\rm in}_{a_{nk}}
\Bigr)
= \frac{1}{e^{\omega^{(-)}_{nk}/kT} -1}.
\end{eqnarray}

Finally, as in the pure electric field case, we find the number of pairs
produced from the initial thermal ensemble
\begin{eqnarray}
n_{nk} (E, B, T) = {\rm Tr} \Bigl(\rho^{\rm in}_{nk} N^{\rm out}_{a_{nk}}
\Bigr) - f_{nk} =
|B_{nk} (E,B) |^2( 2 f_{nk} (T)  + 1).
\label{eb sch}
\end{eqnarray}
In the case of a pure magnetic field, $\alpha_{nk} = 0$ and thereby
$B_{nk} = 0$ as $\Gamma (\alpha_{nk} = 0) = \infty$ and $A_{nk}
= e^{- i \tau \omega_{nk}}$ becomes a pure phase factor.
This confirms the fact that pure magnetic fields cannot produce pairs of
particles and antiparticles even at finite temperature.
The total number of pairs per unit volume and per unit time is given by summing
over the Landau levels and the longitudinal momentum
\begin{eqnarray}
{\cal N} (E, B, T) = \frac{qB}{(2 \pi)^2} \sum_{n} \int dk_z
|B_{nk} (E,B) |^2( 2 f_{nk} (T) + 1).
\end{eqnarray}
Here $(qB)/(2 \pi)$ is the number of Landau levels and another factor $1/(2
\pi)$ is from the $k_z$ integration.

\section{Conclusion}

In this paper we have studied the pair production of charged bosons at finite
temperature by an electric field together with  or without a constant magnetic field.
As an analytically solvable model, the electric field of Sauter-type is used,
which is effectively turned on for a finite period of time.
Upon mode-decomposition the scalar QED Hamiltonian becomes a collection of time-dependent
oscillators. We used the Liouville-von Neumann method to find the exact quantum
states as well as the density operator for
each momentum. The time-dependent gauge potential
for the Sauter-type electric field has two
asymptotic regions where the in-vacuum and out-vacuum are well-defined.
The in-vacuum, which is annulled by the particle
and antiparticle annihilation operators for each momentum
before the onset of the electric field, contains a number of particles
and antiparticles of the same momentum after the electric field is turned on
and thus leads to pair production by the electric field.
We have found that the number of created pairs
for each momentum at the finite temperature is
$n_{nk} (E, B, T) = n_{nk} (E, B, 0) {\rm coth} (\omega^{(-)}_k/2 kT)$,
the number of pairs produced from the vacuum times
a thermal factor from the initial Bose-Einstein distribution.
In the limit of
zero temperature the number of created pairs
reduces to that from the in-vacuum as expected. We also confirm that
the pure magnetic field does not produce pairs even at finite temperature
at one-loop level. We may thus conclude that
the Schwinger pair-production rate is indeed
enhanced by the thermal effect given by the Bose-Einstein distribution.

An interesting problem not handled in this paper is pair production
of fermions at finite temperature by time-dependent electric fields.
The fermion pair production by an arbitrary time-dependent electric
field was studied in Refs. \cite{Dietrich,Cooper-Nayak}. Also the
fermion pair-production rate by the Sauter-type electric field was
found in the WKB instanton action method \cite{Kim-Page07} and in
the worldline instanton method \cite{Dunne-Schubert,DWGS}. However,
to extend fermion pair production to the finite temperature case, we
need a density operator for the time-dependent fermion system. For
instance, the time-dependent annihilation and creation operators in
Ref. \cite{KSK} that satisfy the Liouville-von Neumann equation may
be used to calculate the fermion pair-production rate at finite
temperature. Another interesting problem is the back reaction of the
produced pairs to the initial electromagnetic field. The back
reaction problem in scalar and spinor QED was studied in Ref.
\cite{Cooper-Mottola}. As produced pairs can generate an additional
electromagnetic field to the external field, the produced pairs
provide a source term to the Klein-Gordon equation or the Dirac
equation. The back reaction problem at finite temperature in scalar
and spinor QED will be addressed in a future publication.

\acknowledgments

The authors would like to thank Don N. Page and Misao Sasaki for useful
discussions and comments.
The work of S.~P.~K. was supported by the Korea Science and
Engineering Foundation (KOSEF) grant funded by the Korea government
(MOST) (No. F01-2007-000-10188-0). The work of H.~K.~L was supported
by the Korea Science and Engineering Foundation (KOSEF) grant funded
by the Korea government (MOST) (No. R01-2006-000-10651-0).
\appendix

\end{document}